\documentclass[preprint2]{aastex6}

\usepackage{amsmath,amssymb,amstext}
\usepackage{soul}

\defcitealias{1999ApJ...513..142M}{MLF99}
\defcitealias{2001ApJ...552...91S}{STT01}
\defcitealias{2004ApJ...617.1077F} {F04}
\defcitealias{2011RMxAA..47..113R}{Paper I}
\defcitealias{2013AA...551A..41R}{RH13}

\shorttitle{Metallic winds in dwarf galaxies.}
\shortauthors{Robles-Valdez, F. et al.}

\begin{document}
\title{The metallic winds in dwarf galaxies}
\author{F. Robles-Valdez\altaffilmark{1,2},
  A. Rodr\'iguez-Gonz\'alez\altaffilmark{1},
  L. Hern\'andez-Mart\'inez\altaffilmark{1},
  \and  A. Esquivel\altaffilmark{1}
}

\affil{$^1$ Instituto de Ciencias Nucleares, Universidad
Nacional Aut\'onoma de M\'exico, \\
A.P. 70-543, 04510, Mexico City, Mexico}

\altaffiltext{2}{E-mail: fatima.robles@correo.nucleares.unam.mx}

\begin{abstract}
We present results from models of galactic winds driven by energy
injected from nuclear (at the galactic center) and non-nuclear
starbursts. The total energy of the starburst is provided by
very massive young stellar 
clusters,which can push the galactic interstellar medium and
  produce an important outflow. Such outflow can be a well, or
partially mixed wind, or a highly metallic wind. 
We have performed adiabatic 3D N-Body/Smooth Particle
Hydrodynamics simulations of galactic winds using the {\sc gadget-2}
code. The numerical models cover a wide range of parameters,
  varying the galaxy
concentration index, gas fraction of the galactic disk, and radial
distance of the starburst. We show that an off-center starburst
  in dwarf galaxies is the most effective mechanism to produce a
  significant loss of metals (material from the starburst itself). 
At the same time a non-nuclear starburst produce a high
efficiency of metal loss, in spite of having a moderate to low mass
loss rate.

\end{abstract}


\section{Introduction\label{sec:intro}}
Dwarf galaxies are a crucial ingredient to understand the
  evolution of the large scale universe, as well as  the formation of
  larger galaxies. 
The mass of dwarf galaxies range from $10^7$ to $10^{10}$~M$\odot$
and their luminosities from $10^6$ to $10^{10}$~L$\odot$
\citep{2012AJ....144....4M}
making them the basic units in the formation and evolution of the more massive
galaxies under the $\Lambda$ Cold Dark Matter scenario. However, the
details of how and when this type of galaxies are assembled, and
whether they have galactic winds are issues of current debate 
\citep{1998ARA&A..36..435M,2013AA...551A..41R,2014AdAst2014E...4R}.
Dwarf galaxies are characterised by their poor metallicity and low
masses. They are subdivided in: dwarf irregulars, which have a
significant amount of gas, and thus on-going star formation; dwarf
ellipticals, which are gas-poor with an old stellar population; and dwarf
spheroidal, which are diffuse and practically without gas.
The dwarf irregulars  associated with a burst often develop galactic
winds,  and this can define de metal evolution in this galaxies 
\citep{1997RMxAC...6...36S}. 
Those features can be related to the efficiency
of such galaxies to loss mass and metals by galactic winds, basically 
linking the episodes of star formation or interactions with other galaxies.
The research of these outflows is essential to fully understand
the evolution of dwarf galaxies, their structural properties, and their 
chemical enrichment \citep{2005ARA&A..43..769V}.

The galactic winds, according to their metal content, can be divided
into two general categories \citep{2012ceg..book.....M}: a) ordinary winds, in
which the metals produced and ejected by stars are well mixed with the
gas of the interstellar medium (ISM). Thus the metallicity of the wind
and the ISM are the same.
b) Enriched winds, where the metals produced
and ejected by the stars are not well mixed
with the ISM, i.e. the
chemical elements produced in the supernovae explosions are
carried out to the
intergalactic medium (IGM).
In this case the galactic wind has a metallicity
higher than that of the ISM.
An enriched wind is considered a differential wind
if it has
different ejection efficiencies  of a chemical element to another (e.g. the ejection efficiencies for H and He and for 
heavier elements
(\citealt{1986ApJ...305..669V,2001MNRAS.322..800R,2008A&A...489..555R})).
The efficiency of ejection of each element may differ due
to several factors, most importantly the origin of such chemical
  elements,  since they are produced in different sources that can
  occur at different locations, and times in the ISM. But there
  are also other factors that could affect, such as the medium in which the
  elements mix. For instance, in the case of starbursts, their
  intensity and location (with respect to the potential well of the
  galaxy) become relevant. 
There is a wealth of observational evidence 
of galactic outflows and winds in dwarfs galaxies 
(e.g. \citealt{1988A&A...194...24I,1992AJ....103...60M,1996ApJ...473.1051M,2001ApSSS.277..555P,2007ApJ...671..358W,2007MNRAS.376..523S,2009ApJ...701.1636M,2013A&A...549A.118C}).
Also, recently the
metallicity of outflows in dwarf starburst galaxies has been
determined \citep{2002ApJ...574..663M,2005MNRAS.358.1453O}. At
  the same time, several authors have presented theoretical
  models to explain the properties of the
winds and their impact in the galaxy evolution,
including the chemical analysis of the effect of galactic winds
  in dwarf galaxies
  (\citealt{2010A&A...520A..55Y,2011A&A...531A.136Y}), as well as
  simple chemical models which couple the information of the galactic
  winds in the development of their codes
  (\citealt{1994ApJ...431..598D,1994MNRAS.270...35M,2004MNRAS.351.1338L,2007A&A...468..927L}). There
  have been also hydrodynamic simulations that follow in more detail flows in the
  galaxy   (\citealt{1999MNRAS.309..941D},
  \citealt{1999ApJ...513..142M}, hereafter 
  \citetalias{1999ApJ...513..142M}, \citealt{2001MNRAS.322..800R,2008A&A...489..555R,2011RMxAA..47..113R},
  hereafter \citetalias{2011RMxAA..47..113R}, \citealt{2013AA...551A..41R}, hereafter
\citetalias{2013AA...551A..41R}). Other studies include semi-analytical
simulations that aim to explain the effects of galactic winds at
cosmological scale, and study the interaction of these winds and the
intergalactic medium
\citealt{2007MNRAS.379.1143B,2014MNRAS.445..970D}). These type of
studies usually adopt galactic winds with little detail of their
evolution, focusing on the moment when the material is uncoupled from
the galaxy, and generally do not relate the properties of the wind
with other properties of the galaxy, or they simply assume that
the galactic wind is proportional to the rate of star formation.  
Other authors analyze effects of galactic winds in the intracluster
medium, making large-scale simulations and using observations of
galaxy clusters 
\citep[such as][]{2014A&A...569A..31H}. In these models, efficiencies of
material ejection are averaged given the baryonic
mass of the galaxy. An objective of this paper is the study of
galactic winds in different host galaxies, varying
the properties of the starburst that causes the wind (including the
location of the outbreak in the disk of the galaxy). The results can
provide better estimates for the efficiencies of ejection of gas
mass and metals, thus yielding valuable information to improve the
input for large-scale simulations.

In the present paper we have analyze and compare the efficiencies of
ejection of pristine gas and of metal rich material
in cases where the starburst  occurs at the center and
off-center of the galactic disk, exploring different values of
  the concentration parameter of dwarf galaxies, and  gas fraction
of the galactic disk. 
Our goal is to understand how the ejection of
pristine and enriched gas evolve, and explain the low
metallicity observed in these galaxies, as well as the metal
content in the IGM.   This paper is organized as follows: 
In Section 
\ref{sec:models} we describe the ingredients considered in our galactic
models, the framework to produce the winds is explained in Section
\ref{sec:winds}. 
The results are presented in Section \ref{sec:results} and our
conclusions in Section  \ref{sec:conclusions}.

%
\section{Numerical ingredients of the galaxy}
\label{sec:models}

The density profiles of dark matter halos are not well
constrained for dwarf galaxies, in order to model their 
mass distribution we used  a Hernquist profile
  (Hernquist 1990). 

\begin{equation}
\label{rhoh}
\rho_\mathrm{h}(r)=\frac{M_\mathrm{dm}}{2\pi}\frac{a}{r(r+a)^3},
\end{equation}
where $r$ is the radial distance, and $M_\mathrm{dm}$ is the dark
matter mass. This profile coincides well at the inner parts of the galaxy with
the \citet*[][]{1996ApJ...462..563N} fitting formula, but declines
faster in the outer parts. The total mass converges in the Hernquist
profile, allowing the construction of isolated halos without the need
of an ad-hoc truncation. The $a$ parameter is related to the scale
radius 
 $r_\mathrm{s}$ of the NFW profile by
$a=r_\mathrm{s}\sqrt{2[ln(1+c)]-c/(1+c)}$, where
$c=r_\mathrm{200}/r_\mathrm{s}$ is a `concentration index',
$r_\mathrm{200}$ is the radius  at which the enclosed dark matter mean
density is 200 times the critical  value (where the critical density
is $\rho_\mathrm{crit}=3H^2/8 \pi  G$, see
\citet*[][]{1996ApJ...462..563N}, and \citet{2005MNRAS.361..776S}. We chosen
different concentration indices, c, in the range of the dwarf galaxies,
$5 < c < 15$. The circular velocity of our models is given  by, 
\begin{equation}
\label{vc}
v_\mathrm{c}(r_\mathrm{200})=10\,\mathrm{H_0}\,r_\mathrm{200},
\end{equation}
where H$_\mathrm{0}$ is the Hubble constant.\\

As in \citetalias{2011RMxAA..47..113R} , we modelled gas and star disk components ($M_\mathrm{g}$ and $M_*$, respectively)
with an exponential surface density profile (in the radial direction of the
galactic disk, $R$) with length-scale $R_0$,
\begin{equation}
\label{sigmag}
\Sigma_\mathrm{g}(R)=\frac{M_\mathrm{g}}{2\pi R_0^2}\exp(-R/R_0),
\end{equation}
\begin{equation}
\label{sigmas}
\Sigma_\mathrm{*}(R)=\frac{M_\mathrm{*}}{2\pi R_0^2}\exp(-R/R_0),
\end{equation}

One can obtain the disk mass $M_\mathrm{d}=M_\mathrm{g}+M_*=m_\mathrm{d} M_\mathrm{tot}$), where $m_\mathrm{d}$ is a dimensionless 
parameter (fixed in this work to $m_\mathrm{d}=0.041$), and $M_\mathrm{tot}$ is the total mass of the galaxy, including
the dark matter halo (i.e. $M_\mathrm{tot}=M_\mathrm{d}+M_\mathrm{h}$, where $M_\mathrm{h}$ is the mass
of the halo).\\

The vertical mass distribution of the stars in the disk is specified
by the profile of an isothermal slab with a constant scale 
height $H$. The 3D stellar density in the disk is hence given by,
\begin{equation}
\label{denstar}
\rho_* (R,z)=\frac{M_*}{4 \pi\,H\,R_0^2} {\rm sech}^2\left(\frac{z}{2H}\right)
\exp\left(-\frac{R}{R_0}\right).
\end{equation}

When constructing the galactic models, we assumed that the gas distribution is isothermal and the vertical
structure of the gas disk is in hydrostatic equilibrium:
\begin{equation}
\label{hidros}
\frac{\partial \rho_\mathrm{g}}{\partial z}=-\frac{\rho^2_\mathrm{g}}{ P}
\frac{\partial \Phi_\mathrm{T}}{\partial z},
\end{equation}
where $\Phi_\mathrm{T}=\Phi_\mathrm{h}+\Phi_\mathrm{d}$ is the total gravitational potential 
($\Phi_\mathrm{h}$ and $\Phi_\mathrm{d}$ are the halo and disk gravitational potentials,
respectively).
If one disregards the self-gravity of the gas the above equation
  simplifies, and can be integrated to a closed solution. However, a
  more realistic model requires to include the gravity from all the
  components, and the inclusion of thermal and rotational support,
  which can only be done numerically \citep[for a thorough study
  see][]{2012A&A...543A.129V}. In our models we start with a gaseous
  disk at a temperature of $1000$~K, and an initial guess for the
  density value at the midplane ($z=0$).
For a given  $\Phi_\mathrm{T}$, the solution of this equation is constrained
by the condition
\begin{equation}
\label{sigmag2}
\Sigma_\mathrm{g}(R)=\int \rho_\mathrm{g}(R,z)dz\,,
\end{equation}
where $\Sigma_\mathrm{g}(R)$ is the surface gaseous mass density (see also
Equation~\ref{sigmag}). 
One can obtain the vertical distribution by integrating
Equation~(\ref{hidros}) at a given radius, adjusting the local density in an
iterative process until the desired surface density (Equation~\ref{sigmag})
is recovered. This process is repeated for different radii in order to obtain
an axisymmetric gas density distribution.
With this initial conditions the galaxy models are evolved for an
  additional 10~Myr before the starbursts are imposed.

\subsection{Numerical models}
\begin{figure}[ht!]
\centering
\includegraphics[width=\columnwidth]{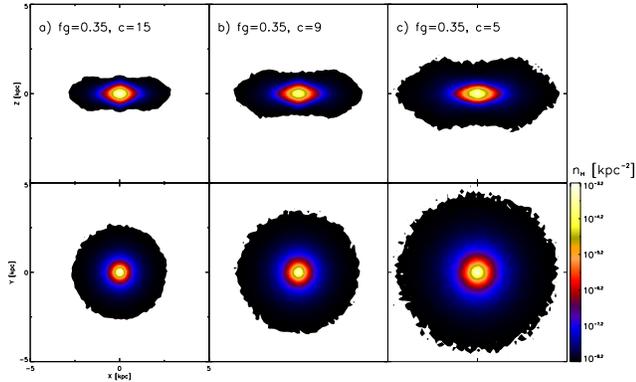}
  \caption{Density projection of the initial condition for the
    G1 galaxy with the same gas fraction, f$_\mathrm{g}$=0.35, and
    with different concentration index: a) $c$=15, b) $c$=9, c) $c$=5. Top
    panels: xz-plane (y=0), taking a slice of the simulation box with
    a width of 0.2kpc. Bottom panels: yz-plane
    (x=0), taking a slice of the simulation box with a width of 0.2kpc.} 
\label{f:fig1}
\end{figure}

We constructed two galaxies: G1 and G2 with a gas
mass of 1.4 $\times$10$^{8}$ M$_\mathrm{\odot}$  and 4.7
$\times$10$^{8}$ M$_\mathrm{\odot}$, respectively, see Table
\ref{tab:models} (they correspond to models G4 and G7 in
\citetalias{2011RMxAA..47..113R}).
We use the Smooth Particles Hydrodynamics (SPH) code
  {\sc gadget-2}  to run a set of adiabatic numerical simulations for dwarf galaxies with the purpose to
study the effects of the: concentration index (c), gas mass fraction
(f$_\mathrm{g}$), the location of the starburst, nuclear and
non-nuclear.
Nuclear starbursts are imposed at the galactic center, while
  non-nuclear starbursts are placed at a galactocentric distance $R$
  along the $x$-axis, but remaining at the mid-plane of the galaxy ($z=0$).
We consider 3 mechanical luminosities (L$_\mathrm{m}$) for the
starburst for each of galaxy. L$_\mathrm{m}$=3, 15 and 75 $\times
10^{39}$ erg~s$^{-1}$ for a host galaxy as G1, L$_\mathrm{m}$=10, 50
and 250 $\times 10^{39}$ erg~s$^{-1}$ for a host galaxy as G2. That
is, the mass of the starbursts are 1.0 $\times$10$^{5}$, 5.0
$\times$10$^{5}$ and 2.5 $\times$10$^{6}$M$_\mathrm{\odot}$ for a
galaxy G1, and 3.3 $\times$10$^{5}$, 1.6 $\times$10$^{6}$  
and 8.3 $\times$10$^{6}$M$_\mathrm{\odot}$ for a galaxy G2, following
a Salpeter initial mass function (Salpeter 1955).
In all the cases, the feedback radius (size of the starbust) is
  $50$\,pc. The energy input by the starburst  is imposed 
 instantaneously, in a single time-step of the simulation
due to the fact that such time-step is of the same order, $1 \times
10^{-5}~\mathrm{Gyr}$ .  The total energy injected was
calculated integrating the L$_\mathrm{m}$ over the cluster lifetime $40~\mathrm{Myr}$, 
(it was described in  \citetalias{2011RMxAA..47..113R}.)
 In \citetalias{2011RMxAA..47..113R} both cases,  adiabatic and
  radiative were considered, and the results showed that (for such
  energetic starbursts) it was easier to lose enriched wind material than
  pristine ISM, regardless of the inclusion of radiative loses. Thus,
  in order to keep the free parameters to a minimum in the present
  paper we only consider the adiabatic case.
In the table we show the different concentration indices
modeled for both galaxies,  with a greater $c$ the galaxy is more compact, and
with a smaller $c$ the galaxy is more disperse, as shown in Figure
\ref{f:fig1}, where we plotted the density projection of the initial
condition with different concentration index for a same gas fraction
(f$_\mathrm{g}$=0.35). This effect of galactic mass distribution is
also reflected in the value of the scale radius of the galaxy
(R$_\mathrm{0}$) for a given $c$,with a decreasing R$_\mathrm{0}$ as $c$
increases (Table \ref{tab:models}). 
We also generated simulations in which $c$ is fixed (taking $c$=9), and
the variable parameter is the gas fraction. The f$_\mathrm{g}$
describes the relative content of gas in the disk, the rest of the
mass is in stars.  The variation of fraction of gas is not reflected in the
R$_\mathrm{0}$ as in the case of c, because the gas content
changes, but the density remains equal. 
Finally, to analyze the impact of the starburst at different distances
from the center of the galaxy (taking $c$=9 and f$_\mathrm{g}$=0.35), we
tested a set of simulations with the starburst located a
galactocentric distance of  R = R$_\mathrm{0}$.
All the numerical simulations in this work were built with three types
of particles, $10^5$ for the disk, $10^5$ for the halo, and $5\times
10^5$ gas particles. This is factor of $\sim5$ higher
resolution than in \citetalias{2011RMxAA..47..113R}. In addition, for
this study, we performed tests with  an order of magnitude less and
more particles (for the three types), finding that although more
detail is seen in the high resolution runs, the efficiencies obtained
do not vary considerably compared with the nominal resolution ones.

\begin{deluxetable}{ccccccc}
\tabletypesize{\scriptsize} 
\tablecaption{The galactic models \label{tab:models}}
\tablecolumns{7}
\tablewidth{0pt}
\tiny{
\tablehead{
\colhead{Galaxy} & \colhead{c} & \colhead{M$_\mathrm{g}$} & 
\colhead{M$_\mathrm{h}$} & 
\colhead{V$_\mathrm{c}$ (r$_\mathrm{200}$)} & 
\colhead{H} & \colhead{R$_\mathrm{0}$} \\ 
\colhead{Model} & \colhead{}  &\colhead{(M$_\odot$)} & \colhead{(M$_\odot$)} &
\colhead{(km s$^{-1}$)} & \colhead{(kpc)}& \colhead{(kpc)}
}
\startdata
G1 & $5 $ & $1.4\times10^{8}$ & $9.4\times10^{9}$ & $34.3$  & $0.118$ & $0.76$ \\
G1 & $9 $  & $1.4\times10^{8}$ & $9.4\times10^{9}$ & $34.3$  & $0.118$ & $0.58$\\
G1 & $15$  & $1.4\times10^{8}$ & $9.4\times10^{9}$ & $34.3$  & $0.118$ & $0.45$\\
G2 & $5 $  & $4.7\times10^{8}$ & $3.2\times10^{10}$ & $51.9$ & $0.178$  & $1.16$\\
G2 & $9 $ & $4.7\times10^{8}$ & $3.2\times10^{10}$ & $51.9$ & $0.178$  & $0.88$ \\
G2 & $15$  & $4.7\times10^{8}$ & $3.2\times10^{10}$ & $51.9$ & $0.178$  & $0.68$
\enddata
}
\end{deluxetable}

\section{Metal rich galactic winds.}\label{sec:winds}

The most used outflow rate is the well mixed galactic winds, it
represent the case where the mixing between the winds and the ISM is
  efficient. This result in a uniform metallicity in the galactic wind
  and the host galaxy \cite[][and references therein]{2012ceg..book.....M}.
The amount of gas and chemical species as a function of the lifetime
of an isolated galaxy, considering instantaneous recycling
approximation \citep{2012ceg..book.....M} are given by,

\begin{equation}
\label{eq:M$_{g}$}
\frac{dM_\mathrm{gas}(t)}{dt}=-(1-R) \psi(t)+f(t) - w(t),
\end{equation}

and

\begin{eqnarray}
\label{eq:zM$_{g}$}
\frac{d [Z_{i}(t) M_\mathrm{gas}(t)]}{dt} 
& = & -(1-R) \psi(t) Z_i(t)  \nonumber \\
 &  &+(1-R_\mathrm{ms})\psi(t) y_{i,\mathrm{ms}}\nonumber\\ 
 & &+(1-R_\mathrm{lims})\psi(t) y_{i,\mathrm{lims}}\nonumber\\
& &+Z^\mathrm{f}_{i} f(t)- Z^\mathrm{w}_{i}(t) w(t)\,,
\end{eqnarray}
where $M_\mathrm{gas}$ is the gas mass, $w(t)$ the mass
outflow from the galactic potential, $\psi(t)$ is the star
formation rate, $Z_{i}(t)$ is the time dependent ISM metallicity
for the $ith$ source,
$y_{i,\mathrm{ms}}$ is the corresponding stellar yield from
massive stars ($=\int^{M_\mathrm{up}}_\mathrm{8 {\rm M}_\odot}m p_{i}(m) \phi(m)
dm/(1-R_\mathrm{ms})$),
$y_{i,\mathrm{lims}}$ is the stellar yield of low and intermediate mass
stars $[=\int^{8 {\rm M}_\odot}_\mathrm{M_\mathrm{low}}m p_{i}(m)
\phi(m) dm/(1-R_\mathrm{lims})]$. The mass fraction returned to
the ISM is $R_\mathrm{ms}=\int^{M_\mathrm{up}}_\mathrm{8 {\rm
    M}_\odot}(m-m_r) \phi(m) dm$, while
$R_\mathrm{lims}=\int^{8 {\rm M}_\odot}_\mathrm{M_\mathrm{low}}(m-m_r)
\phi(m) dm$. $R$ is the sum of $R_\mathrm{ms}$ and
$R_\mathrm{lims}$. $M_\mathrm{low}$ and $M_\mathrm{up}$ are the lower
and upper limits of the initial mass function, $\phi(m)$.   We
  used M$_\mathrm{low}=1~M_\odot$ and M$_\mathrm{up}=100~M_\odot$ for
the mechanical energy calculations.

The term $ Z^\mathrm{w}_{i}(t) w(t) $ of Equation~\ref{eq:zM$_{g}$} 
assumes that the interstellar medium of the host galaxy is well
mixed, at least locally  (in the starburst region).

In the starburst region the superwind
contains the metals produced and ejected by the stars via
supernova (SN) explosions, and stellar winds (in main and post-main
sequences). 

There are several mechanisms that came into play to produce, or
  inhibit the production of a well mixed wind. For instance,
 the process of mixing of the metals (formed in
the star formation event) with the interstellar medium of the 
galaxy requires a very long diffusion time. 
Such diffusion time is  around  10$^{5}$ and 10$^{14}$ yr for
the ionised and molecular regions, respectively
\citep[see][and references therein]{1996AJ....111.1641T}.
\citet{1996AJ....111.1641T} presented a model  
based on the galactic fountains that are able to push hot gas
injected by the massive stars from the galactic disk out to the galactic
halo. In his model the galactic fountain is made up of metal rich clumps
that travel through the halo of the galaxy and fall back to
different regions of the galactic disk. As a result the metals
are well distributed in the galaxy, mixing with the local interstellar
medium and enhancing the process of contamination.
However, as pointed out by \cite{2001MNRAS.322..800R}, it is
  necessary to consider that in these models the cooling of the rich
  clumps could be very important to establish the time in which the
  metals diffuse, given that it affects its  fall back the galactic
  disk.
In other words, the process of mixing is dependent of the cooling by
metals and hence the heating efficiency adopted for SNs also it should
be taken into account.
Additionally, tidal stripping can produce outflows that are
inherently well mixed. At the same time, a star formation event with
only a few massive stars can increase the metallicity locally, without
being able to affect the ISM in the entire host galaxy.
 
\citetalias{1999ApJ...513..142M} and several more recent papers
(e.g. \citetalias{2011RMxAA..47..113R}) study the efficiency at which
metal rich material is lost by dwarf galaxies
with an important star formation event. The 
general conclusion is that the efficiency of ejection of the
  metal rich material from the starburst ({\it metal injection
    efficiency} in short, $\xi_\mathrm{z}$) is not
   tied to the total mass ejection
efficiency ($\xi_\mathrm{m}$).
As a result, many models produce a partially mixed galactic
  wind, in which both the metals produced in the wind, and the
  pristine ISM material contribute to the galactic outflow.

Using the same definitions of \citetalias{2011RMxAA..47..113R}, the
mass ejection efficiency is given by
\begin{equation}
\label{eq:w}
\xi_\mathrm{m}=\frac{M_\mathrm{ej}}{M_\mathrm{gas}},
\end{equation}
\noindent
where M$_\mathrm{ej}$ is the gas mass of the galaxy ejected (unbound)
and  M$_\mathrm{gas}$ is the total gas mass. 

And the metal ejection efficiency is

\begin{equation}
\label{eq:w1}
\xi_\mathrm{z}=\frac{M_\mathrm{c,ej}}{M_\mathrm{c}},
\end{equation}
\noindent
where M$_\mathrm{c}$ is the total mass injected via stellar winds and
supernova remnants inside the radius of feedback, R$_{c}$,  
and M$_\mathrm{c,ej}$ is the mass ejected (unbound) that
originated in the starburst. In the present paper,  
To obtain  M$_\mathrm{ej}$ from the simulations,
we consider their position and velocity. If their velocity is greater
than the escape velocity at their location; and if they lie outside a
cylindrical region of radius $20$\,kpc and $40$\,kpc in height, at the
center of the galaxy ; they are considered as mass that is effectively
lost form the galaxy \citep[see also][]{1999MNRAS.309..941D}.

Assuming a thermal wind driven by an instantaneous
starburst event, the galactic wind is a function of the mass and metal
ejection efficiencies at the time that the starburst occurs. We
propose to use the average  mass lost rate as function of mass and
metal efficiency as:
\begin{eqnarray}
\label{eq:zM$_{g}$mix}
w(t)=
\begin{cases}  
 &\rm{if}  \;\;t \;\; \rm{is~not~within~~} \Delta t_\mathrm{ lf}~:\\
 & 0, \\
 & \\
 &\rm{if}  \;\;t \;\; \rm{is~within~~} \Delta t_\mathrm{lf}~:\\
 &\xi_\mathrm{m}  \frac{M_\mathrm{gas}}{\Delta t_\mathrm{lf}}+\xi_\mathrm{z} \frac{M_\mathrm{c}}{\Delta t_\mathrm{lf}}, \\
\end{cases}
\end{eqnarray}
where $\Delta t_\mathrm{lf}$ is the lifetime of the massive
stars, which is around $10^7$ yr). 

This way, one can modify the chemical
evolution expression in Equation~\ref{eq:zM$_{g}$} to include
the metal ejection efficiency in favour of the mass outflow function
$w(t)$. For instance, we can replace the $Z_\mathrm{i}^\mathrm{w}w(t)$
term in Equation \ref{eq:zM$_{g}$}, for
{$(1-R_\mathrm{ms})\,\xi_\mathrm{z}\,\psi(t) y_{i,\mathrm{ms}}$} if
$t$ is within $\Delta t$.

\section{Previous works}\label{sec:works}

\citetalias{1999ApJ...513..142M} studied the effect of a starburst
placed at the center of dwarf galaxies with gas masses between
10$^{6}$ - 10$^{9}$M$_{\odot}$.  
They found that the galaxies with masses of 10$^{6}$M$_{\odot}$ have
a high efficiency of gas ejection, which is
  mostly independent from the energy of the starburst 
(mechanical luminosity, L$_\mathrm{m}$). If the galaxy mass is $\leq$
10$^{8}$M$_{\odot}$  the efficiency of gas ejection is
relatively low and still dependent on L$_\mathrm{m}$, while for
10$^{9}$M$_{\odot}$ the efficiency is almost null. In the case
of the metal ejection efficiency (the efficiency to
  loose the metal rich material from the starburst) for galaxies with
{$\leq$10$^{8}$M$_{\odot}$} is close to unity and does not
depend on L$_\mathrm{m}$,  
whereas for 10$^{9}$M$_{\odot}$ depends significantly
on L$_m$. That is, the metals produced in the starburst escape
more easily from the galaxy that the pristine gas (H and He,
see tables 2 and 3 of \citetalias{1999ApJ...513..142M}).

\citetalias{2001ApJ...552...91S} consider the
same range of gas mass of the galaxies that
\citetalias{1999ApJ...513..142M}. They studied the expulsion of gas
and its relation to the degree of flatness/roundness of the
galaxy, and whether the ejection is in the same direction as its axis of
rotation. They also take into account the pressure of the IGM and the
existence of gaseous halos. They conclude that the super-bubbles in
flat galaxies with rotation, tend to break out from the disk
  and expel their metal content with only a small amount of
pristine ISM gas (the cases of 
\citetalias{1999ApJ...513..142M}, fall into these type of galaxies,
with low pressure of the IGM). To eject the contents of 
super-bubbles in spherical galaxies a starburst 3 times more
energetic is required in comparison to a flat galaxy (see
figure 3 of \citetalias{2001ApJ...552...91S}). 

\citet{2004ApJ...617.1077F} (hereafter
\citetalias{2004ApJ...617.1077F}) modeled disk galaxies with masses of
10$^{9}$M$_{\odot}$ and one of 10$^{8}$M$_{\odot}$, with discrete
events of supernovae (SNe) distributed over a fraction of the galactic
disk (and one off-centered model).
Comparing their models 1 and 2 (same parameters, but in model 1 the SN
explosions are all placed at the galactic center, while in model 2 all
are placed at the same point off-center). They found that the metal
efficiency ejection is higher when  the SNe are concentrated in the
center (see Section \ref{sec:results} for a comparison with our
results). 

In \citetalias{2011RMxAA..47..113R}, a wide range of galactic masses
 was explored (M$_\mathrm{g}$ = 6$\times$10$^{6}$ -
10$^{11}$M$_{\odot}$) and central starburst masses (M$_\mathrm{SB}$ =
10$^{2}$ - 10$^{7}$M$_{\odot}$). The results are in good agreement
with those of previous studies, metal efficiency ejection and gas
decrease as the mass of the galaxy increases. It is also shown that the
values of efficiencies are higher for the more massive bursts (high
L$_m$). Due to the larger mass range of galaxies and starbursts, a
model that makes the transition between regimes of low and
high efficiencies of ejection was included: M$_\mathrm{SB}$ =
3$\times$10$^{4}$M$_{\odot}$ 
and M$_\mathrm{g}$ = 3$\times$10$^{7}$M$_{\odot}$ (for more details see
figures 2 to 5 of \citetalias{2011RMxAA..47..113R}).
In \citetalias{2011RMxAA..47..113R} 
the variation in location of of the
starburst for different models was considered, finding that
the relation of the ejection efficiency with distance from the 
radius of the starburst is not monotonic.
For the most massive galaxies the
fraction of unbound pristine gas was less than 25 percent, while the
less massive galaxies lose virtually all their gas content. For the
enriched material a large escape fraction is seen in the less massive 
galaxies, in contrast with the less energetic starbursts, in
  which most of the metal content is retained, the results
are qualitatively consistent with those obtained by
\citetalias{1999ApJ...513..142M}. 

\citetalias{2013AA...551A..41R} made a similar study to
\citetalias{2001ApJ...552...91S} for galaxies with masses between
10$^{7}$ - 10$^{9}$M$_{\odot}$, analyzing the ejection efficiencies
varying the mass distribution of the host galaxy.
 As \citetalias{2001ApJ...552...91S}, they concluded that the 
ejection of enriched gas on a flat galaxy is higher than on a
spherical, while the fate of the pristine gas is
relatively independent of the geometry of the galaxy. 
They also confirm that metal and gas efficiencies
ejection are strongly dependent on the galactic mass. Thus, smaller
galaxies develop larger flows and the fraction of metals and gas
ejected tend to be larger (see table 2 of
\citetalias{2013AA...551A..41R}). In addition,
they found  that the fate of 
pristine gas and of newly produced metals are strongly
dependent on the mass of the galaxy.

\section{Results and discussion}\label{sec:results}

We computed the fraction of total mass and metal rich material
(produced in the starburst) that is effectively lost in all the
models. After the starburst occurs, the mass and metal ejection efficiencies
increase, and after a few tenths of $\mathrm{Gyr}$ they saturate.

In what follows we discuss the
effects of the concentration index, the disk gas fraction, and the
location of the starburst. The efficiencies were calculated in all
cases after an integration time of $t=1~\mathrm{Gyr}$
{\bf(i.e. $990~\mathrm{Myr}$} after the injection of the
  starburst).

\subsection{The concentration index effects}

\begin{figure*}[!ht]
\centering
\includegraphics[width=1.8\columnwidth]{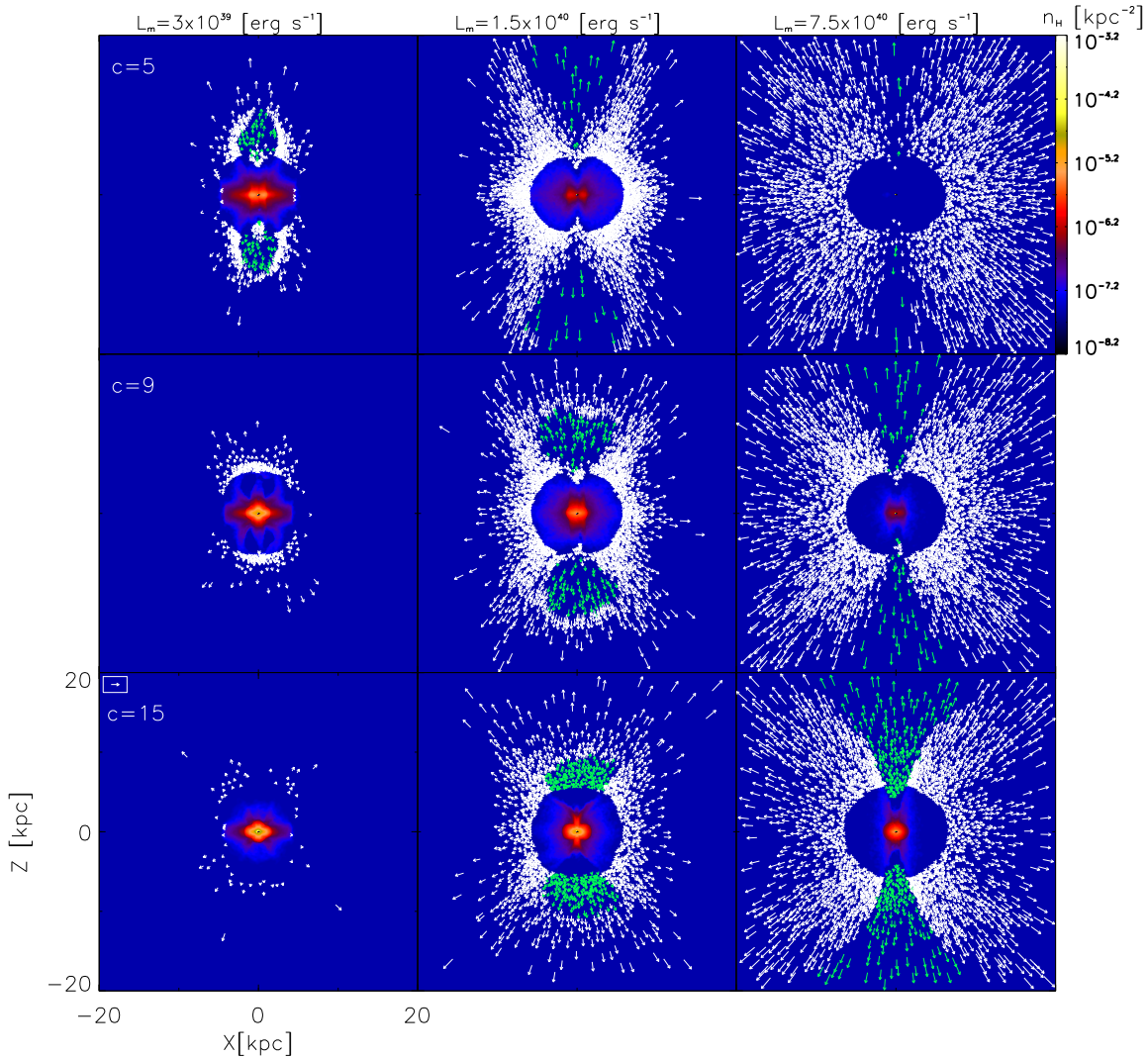}
  \caption{Density cut of the G1 galaxy, xz-plane (y=0), at 
    $t=0.1~\mathrm{Gyr}$ (the slice has a thickness
      of $0.2~\mathrm{kpc}$). The star formation event is
    nuclear (at R=0). The arrows depict the
    velocity field of particles unbound from the potential
    well of the galaxy, the white arrows represent the pristine gas mass and
    the green ones the metals, for visual purposes we plotted
    only 20\% of the arrows. In this figure we show simulations with the same
    gas fraction, f$_\mathrm{g}$=0.35, with different
    concentration index, top row: $c$=5, middle row: $c$=9, bottom
    row: $c$=15; and mechanical luminosities, left
    column: L$_\mathrm{m}$=3, 
    middle column: L$_\mathrm{m}$=15, and right
    column: L$_\mathrm{m}$=75 (units: 
    $\times 10^{39}$ erg s$^{-1}$).}
\label{fig2}
\end{figure*}

Figure \ref{fig2} shows the effect of the concentration index in
the G1 galaxy after an integration time of $t=0.1~\mathrm{Gyr}$ (the
starburst occurs at  $t=0.01~\mathrm{Gyr}$, at the center of the galaxy).
In this Figure, it can be noted
that more ISM mass (white arrows), and also more metals (green arrows)
escape in the models where the concentration of the galaxy is smaller
(upper panels). This is due to the fact that the galaxy is less dense
for a smaller $c$, and thus it is easier for the energy produced by
supernova explosions to break out the bubbles, emptying their content
(metal rich) into the IGM, carrying in some cases a significant
fraction of pristine ISM along.

The least efficient model in mass and metals loss it is that with a
larger $c$ and a smaller L$_\mathrm{m}$ (bottom left panel), and the most
efficient has a smaller $c$ and a larger L$_\mathrm{m}$ (upper right
panel). 
This can be confirmed in the top panels of 
Tables \ref{tab:emr0} and \ref{tab:ezr0}, where we list the mass and
metals ejection efficiencies for all the models where the starburst
occur at the center of the galaxy.

In Table \ref{tab:emr0} (top) we show the values of the mass
ejection efficiency as a function of $c$ and L$_\mathrm{m}$ for both
galaxies, with a nuclear starburst. As described for G1 in Figure \ref{fig2},
it is noted that with a lower $c$ and a greater L$_\mathrm{m}$, higher
efficiencies are achieved. The G2 galaxy has a mass ejection
efficiency low with respect to G1.
The reason is that despite that G2 has more energetic
starbursts, it is also is more massive, and does not allow at the
super-bubbles to break as easily as in G1.
Similarly, in Table 3 (top box) we present the
values of metal ejection efficiencies, where it can be seen
the same trend that that the models with smaller $c$ and
larger L$_\mathrm{m}$ are very efficient to launch the metals out of
the galaxy. 

\begin{figure*}[!ht]
\centering
\includegraphics[width=1.8\columnwidth]{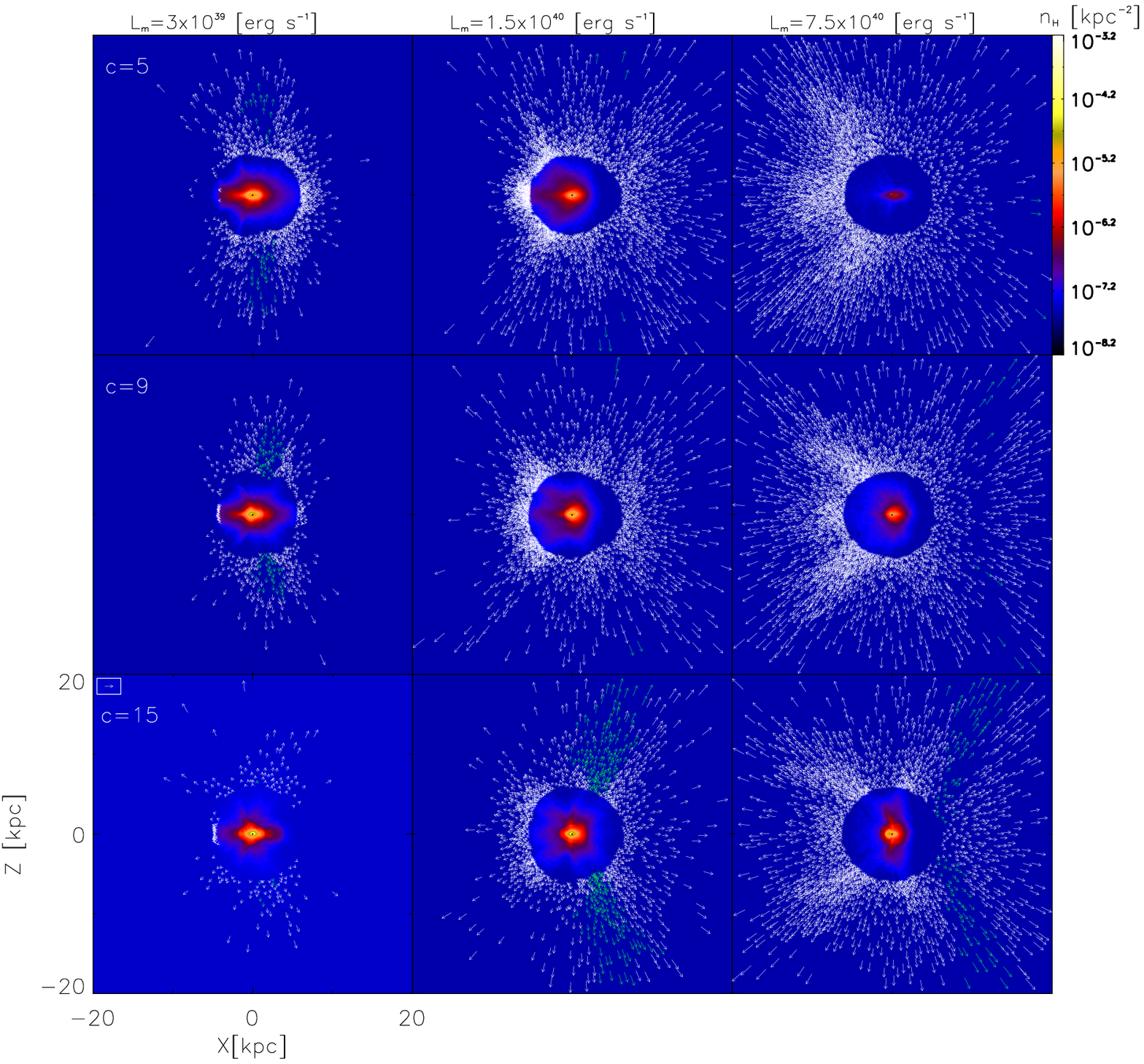}
  \caption{Same as in Figure \ref{fig2}, but for a
    non-nuclear a star formation event (at R=R$_\mathrm{0}$).} 
\label{fig3}
\end{figure*}

In Figure \ref{fig3} we show the G1 galaxy with the same
parameters of Figure \ref{fig2}, only that in this case the
starburst is non-nuclear off the galactic center, and the
values of the mass ejection efficiency and metal ejection efficiency
are in the Tables \ref{tab:emr0} to \ref{tab:ezrH} (top boxes),
respectively. If we compare the efficiencies of loss 
of mass and of metals between nuclear and non-nuclear
starbursts, we can see that there is not much difference, although in 
general it is easier to lose mass gas in the models with non-nuclear
starburst, the metal loosing efficiencies are rather similar.

\begin{figure*}[!ht]
\centering
\includegraphics[width=1.8\columnwidth]{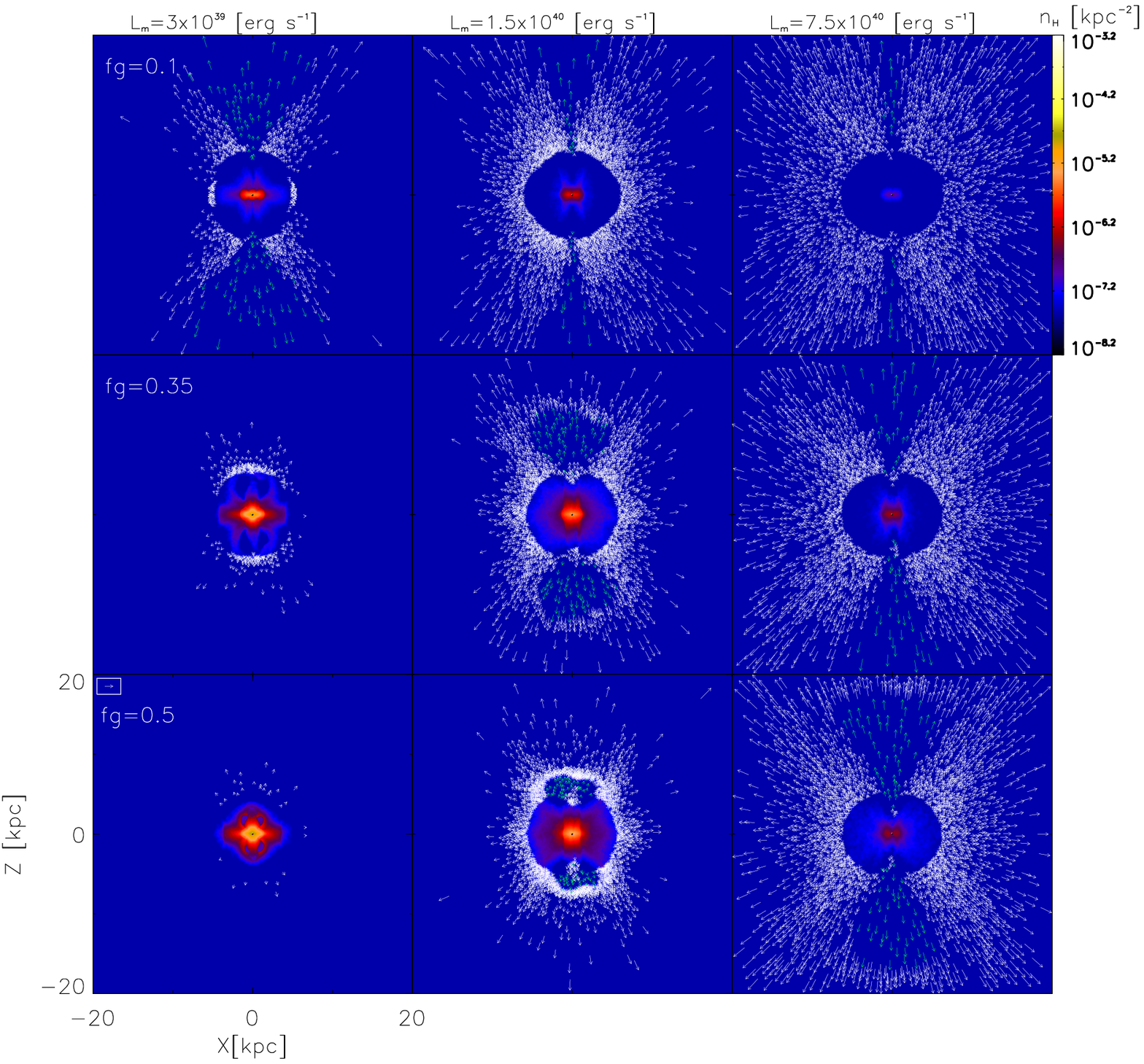}
  \caption{Density cut of the G1 galaxy, xz-plane (y=0), at
    $t=0.1~\mathrm{Gyr}$ (the slice has a thickness
      of $0.2~\mathrm{kpc}$). The star formation event is nuclear
    (at R=0). The arrows edit1{depict} the
    velocity field of particles unbound from the potential
    well of the galaxy, the white arrows represent the pristine gas mass and
    the green ones the metals, for visual purposes we plotted
    only 20\% of the arrows.
    In this figure we show the simulations with different gas
    fraction, top row: f$_\mathrm{g}$=0.1, middle row:
    f$_\mathrm{g}$=0.35, bottom row: f$_\mathrm{g}$=0.5 and with
    the same concentration index, $c$=9; vs mechanical luminosity,
    left column: 
    L$_\mathrm{m}$=3, middle column: L$_\mathrm{m}$=15, and
    right column: L$_\mathrm{m}$=75 (units: $\times 10^{39}$ erg s$^{-1}$).} 
\label{fig4}
\end{figure*}

\begin{figure*}[!ht]
\centering
\includegraphics[width=1.8\columnwidth]{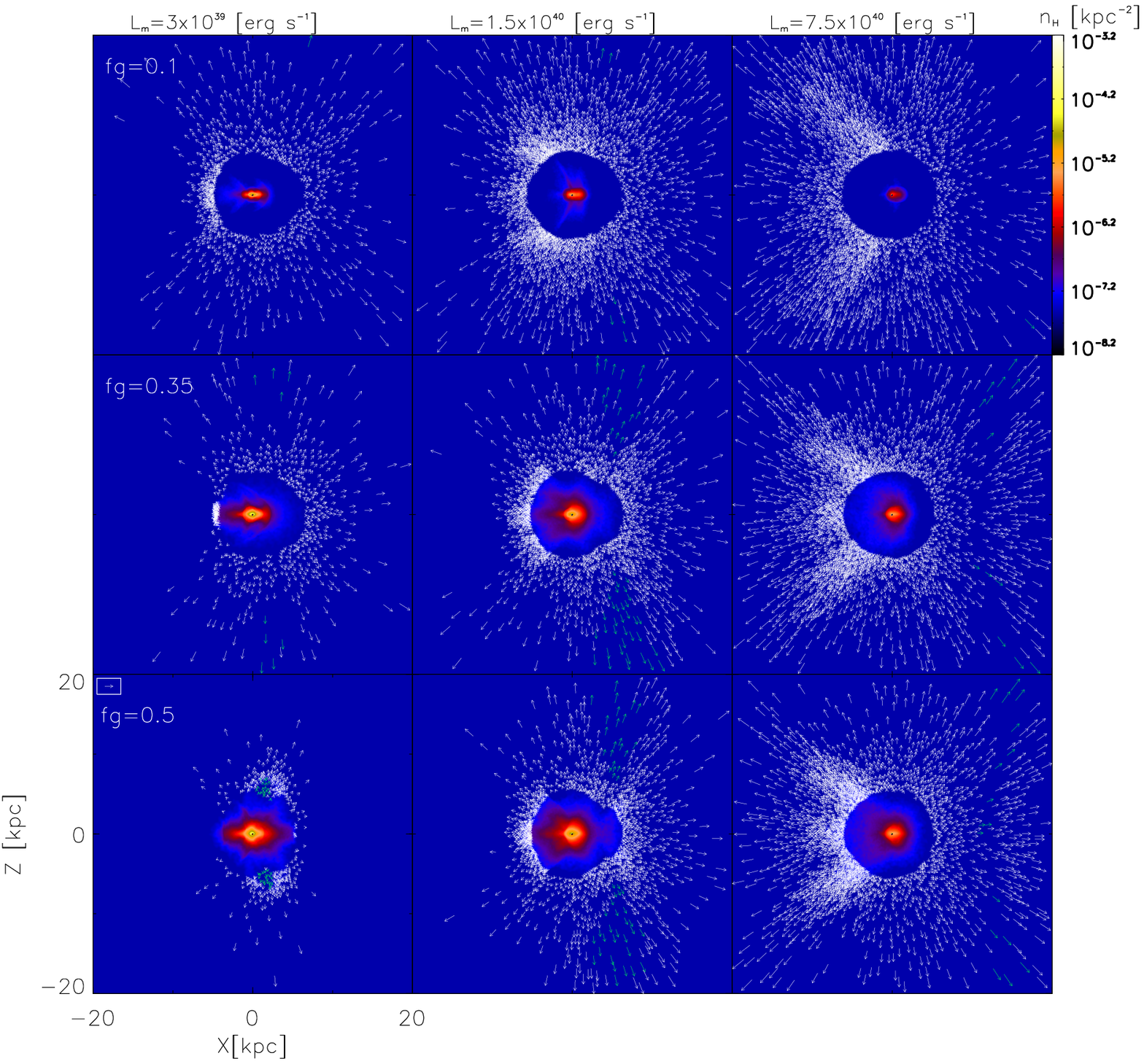}
  \caption{Same as Figure \ref{fig4}, but for a non-nuclear star
    formation event (at R=R$_\mathrm{0}$).} 
\label{fig5}
\end{figure*}

\begin{table}[]
\caption{Mass ejection efficiency of nuclear starbursts}
\begin{center}
\begin{tabular}{ c | c c c | c c c}
\hline
\hline
  &   \multicolumn{3}{c |}{L$_\mathrm{m}$ ($\times 10^{39}$ erg
    s$^{-1}$)} 
  & \multicolumn{3}{c}{L$_\mathrm{m}$ ($\times 10^{39}$
                 erg s$^{-1}$)}   \\\hline
  & $3$ & $15$ & $75$ & $10$ & $50$ & $250$  \\\hline
c  &  \multicolumn{3}{c |}{G1 ($f_\mathrm{g}=0.35$)
  }&  \multicolumn{3}{c  }{G2 ($f_\mathrm{g}=0.35$)} \\
 \hline
 \hline
 $ 5$ &  $0.04  $  & $0.36$ & $0.90$ & $0.01  $ & $0.10$ & $0.59$ \\
 $ 9$ &  $< 0.01$  & $0.23$ & $0.72$ & $< 0.01$ & $0.06$ & $0.29$ \\
 $15$ &  $< 0.01$  & $0.14$ & $0.58$ & $< 0.01$ & $0.04$ & $0.19$ \\
\hline
  f$_\mathrm{g}$ &  \multicolumn{3}{c |}{G1 ($c=9$)}
               &  \multicolumn{3}{c  }{G2 ($c=9$)} \\  
  \hline
  \hline
  $0.1 $  & $0.09  $   & $0.38$ & $0.85$  & $ 0.03 $ & $0.10$ & $0.31$  \\
  $0.35$  & $< 0.01$   & $0.23$ & $0.72$  & $< 0.01$ & $0.06$ & $0.29$  \\
  $0.5 $  & $< 0.01$   & $0.17$ & $0.69$  & $< 0.01$ & $0.04$ & $0.28$  \\
\hline 
\hline
\end{tabular}
\end{center}
\label{tab:emr0}
\end{table}%

\begin{table}[]
\caption{Metal ejection efficiency of nuclear starbursts}
\begin{center}
\begin{tabular}{c | c c c | c c c}
\hline
\hline
  &   \multicolumn{3}{c |}{L$_\mathrm{m}$ ($\times 10^{39}$ erg
    s$^{-1}$)} 
  & \multicolumn{3}{c}{L$_\mathrm{m}$ ($\times 10^{39}$
                 erg s$^{-1}$)}   \\\hline
  & $3$ & $15$ & $75$ & $10$ & $50$ & $250$  \\\hline
$c$  &  \multicolumn{3}{c |}{G1 ($f_\mathrm{g}=0.35$)
  }&  \multicolumn{3}{c  }{G2 ($f_\mathrm{g}=0.35$)} \\
 \hline
 \hline
 $ 5$  & $0.82$ & $1.0$  & $1.0 $ & $0.49 $ & $1.0 $ & $1.0 $ \\
 $ 9$  & $0.0 $ & $1.0$  & $1.0 $ & $<0.01$ & $0.98$ & $1.0 $ \\
 $15$  & $0.0 $ & $0.30$ & $0.93$ & $<0.01$ & $0.42$ & $0.98$ \\
\hline
  f$_\mathrm{g}$
  &  \multicolumn{3}{c |}{G1 ($c=9$)}
  &  \multicolumn{3}{c  }{G2 ($c=9$)} \\  
 \hline
 \hline
  $0.1 $ & $0.99$ & $1.0 $ & $1.0$  & $0.99  $  & $1.0 $ & $1.0$  \\
  $0.35$ & $ 0.0$ & $1.0 $ & $1.0$  & $< 0.01$  & $0.98$ & $1.0$  \\
  $0.5 $ & $ 0.0$ & $0.06$ & $1.0$  & $<0.01 $  & $0.83$ & $1.0$  \\
\hline 
\hline
\end{tabular}
\end{center}
\label{tab:ezr0}
\end{table}%

\subsection{The effect of the disk gas fraction}

Figure \ref{fig4} illustrates the effect of the disk gas fraction
in the G1 galaxy, at $t=0.1~\mathrm{Gyr}$, also for a nuclear 
starburst.
The loss of gas mass and of enriched gas decrease from
upper panels to the lower panels. That is, for models with larger
f$_\mathrm{g}$ it is more difficult for the bubble to break out of
  the disk because there is a larger amount of gas in the disk to
be cleared. 
Similarly to models with variable $c$, a higher L$_\mathrm{m}$
makes it easier to break out of the disk, and therefore as the
mechanical luminosity increases the amount lost of gas mass and metals
does as well.
Then, the least efficient model in loss of mass and
metals have a larger fraction of gas combined with a lower L$_\mathrm{m}$
(bottom left panel), and the most efficient have a smaller
f$_\mathrm{g}$  and large L$_\mathrm{m}$ (upper right panel). 

In Table \ref{tab:emr0} (bottom panels) we show the values of the mass
ejection efficiency as a function of f$_\mathrm{g}$ and L$_\mathrm{m}
$ for both galaxies, with the nuclear starburst. Not
  surprisingly the G2 galaxy results in efficiencies lower
that the those of G1, since it is more  massive than
G1, having a larger amount of gas mass to push out of the
galaxy by the supernova explosions. We show in 
Table \ref{tab:ezr0} (bottom box) 
the metal ejection efficiencies for both galaxies, and can see
again that the models with smaller f$_\mathrm{g}$ and
  L$_\mathrm{m}$ large are more efficient to clear the metals
  from the galaxy.  

Figure \ref{fig5} shows the G1 galaxy with the same parameters of
Figure \ref{fig4}, but with non-nuclear starburst, and
the values of the mass ejection efficiency and metal ejection
efficiency  are in the Tables  \ref{tab:emrH} and \ref{tab:ezrH}
(bottom boxes), respectively.

\subsection{The non-nuclear starburst effect}

\begin{figure}[!ht]
\centering
\includegraphics[width=\columnwidth]{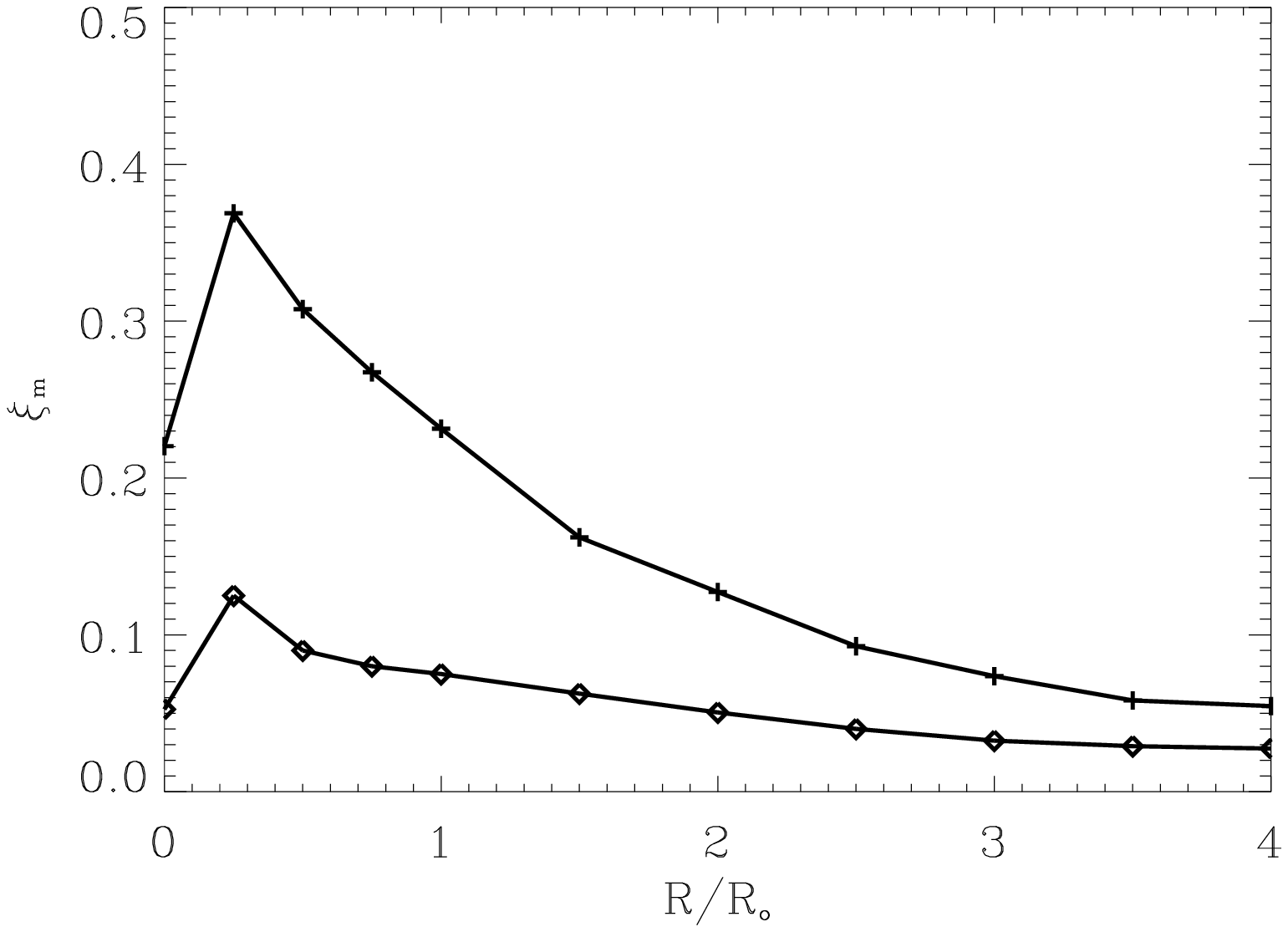}
  \caption{Gas (galactic) mass ejection efficiency
    $\xi_\mathrm{m}$ computed for G1 and G2 model
    galaxies with gas fraction f$_\mathrm{g}$=0.35, concentration
    index $c$=9, and 
    mechanical luminosity of L$_\mathrm{m}$=15 for G1 (solid line with
    crosses) and L$_\mathrm{m}$=50 for G2 (solid line with diamonds),
    as a function of the cylindrical radius R/R$_\mathrm{0}$ at which
    the star formation event is imposed.}  
\label{fig6}
\end{figure}

In Figure \ref{fig6} we show the gas mass ejection
efficiency for the G1 and G2  galaxies with $c$=9 and
f$_\mathrm{g}$=0.35, for different values of $R$
(namely, varying the position of the starburst only on the x axis). 

The general behavior of $\xi_\mathrm{m}$ for both
galaxies is similar, but the values differ since G1 has larger
efficiencies than G2.  As mentioned before this is because G1 is
less massive than G2, and then G1 has a weaker gravitational potential
overcome, therefore the wind is produced considerably easier in
G1.
In the central part of the galaxy the ejection is less
effective, since the potential in this zone is stronger. Out of the
central zone the potential is weakened and it is easier for the
material to escape, therefore a peak in the behavior of the
efficiencies can be observed. 
At larger radii however, in spite of being easier to launch
  galactic material out of the potential well, there is less mass in
  the outskirts of the galaxy to do so.

\begin{figure}[!ht]
\centering
\includegraphics[width=\columnwidth]{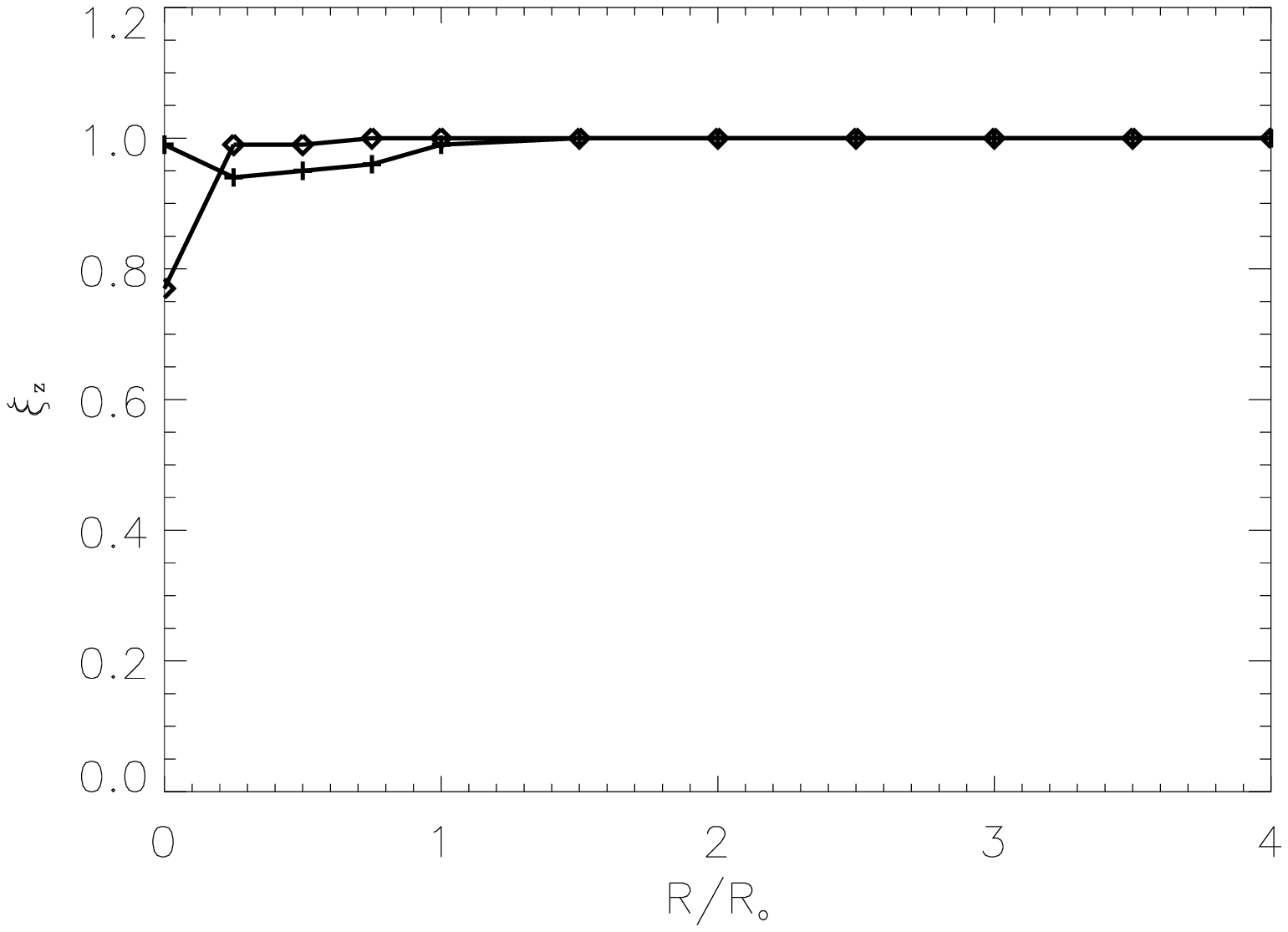}
  \caption{Metal ejection efficiency $\xi_\mathrm{z}$ computed for G1
    and G2 model galaxies with gas fraction
    f$_\mathrm{g}$=0.35, concentration index $c$=9, and mechanical
    luminosity of L$_\mathrm{m}$=15 for G1 (solid line with diamonds)
    and L$_\mathrm{m}$=50 for G2 (solid line with crosses), as a
    function of the radius cylindrical R/R$_\mathrm{0}$ at which
    the star formation event is imposed.}  
\label{fig7}
\end{figure}

With the same conditions of Figure \ref{fig6},
we show in Figure \ref{fig7} the metal ejection efficiency 
for nuclear and non-nuclear starbursts. It is observed that the loss of
metals is 100 percent for both galaxies models for
all of starbursts placed beyond one scale-length. This is due mostly
to the fact that at large $R$ there is not enough  mass to slow down
the metal rich injection form the starburst.

\begin{table}[]
\caption{Mass ejection efficiency of non-nuclear starbursts (at R=R$_\mathrm{0}$)}
\begin{center}
\begin{tabular}{ c | c c c | c c c}
\hline
\hline
  &   \multicolumn{3}{c |}{L$_\mathrm{m}$ ($\times 10^{39}$ erg
    s$^{-1}$)} 
  & \multicolumn{3}{c}{L$_\mathrm{m}$ ($\times 10^{39}$
                 erg s$^{-1}$)}   \\\hline
  & $3$ & $15$ & $75$ & $10$ & $50$ & $250$  \\\hline
$c$  &  \multicolumn{3}{c |}{G1 ($f_\mathrm{g}=0.35$)
  }&  \multicolumn{3}{c  }{G2 ($f_\mathrm{g}=0.35$)} \\
 \hline
 \hline
 $ 5$ & $0.07$ & $0.33$ & $0.85$ & $0.02$ & $0.13$ & $0.36$  \\
 $ 9$ & $0.03$ & $0.28$ & $0.61$ & $0.00$ & $0.06$ & $0.27$  \\
 $15$ & $0.01$ & $0.14$ & $0.47$ & $0.00$ & $0.04$ & $0.18$  \\
\hline
  f$_\mathrm{g}$ 
&  \multicolumn{3}{c |}{G1 ($c=9$)}
&  \multicolumn{3}{c  }{G2 ($c=9$)} \\ 
  \hline
  \hline
 $0.1 $ & $0.12$ & $0.36$ & $0.74$ & $0.03$ & $0.12$ & $0.35$   \\
 $0.35$ & $0.03$ & $0.28$ & $0.61$ & $0.00$ & $0.06$ & $0.27$  \\
 $0.5 $ & $0.01$ & $0.16$ & $0.57$ & $0.00$ & $0.05$ & $0.24$  \\
\hline 
\hline
\end{tabular}
\end{center}
\label{tab:emrH}
\end{table}%

\begin{table}[]
\caption{Metal ejection efficiency of non-nuclear starbursts (at R=R$_\mathrm{0}$)}
\begin{center}
\begin{tabular}{c | c c c | c c c}
\hline
\hline
  &   \multicolumn{3}{c |}{L$_\mathrm{m}$ ($\times 10^{39}$ erg
    s$^{-1}$)} 
  & \multicolumn{3}{c}{L$_\mathrm{m}$ ($\times 10^{39}$
                 erg s$^{-1}$)}   \\\hline
  & $3$ & $15$ & $75$ & $10$ & $50$ & $250$  \\\hline
$c$  &  \multicolumn{3}{c |}{G1 ($f_\mathrm{g}=0.35$)
  }&  \multicolumn{3}{c  }{G2 ($f_\mathrm{g}=0.35$)} \\
 \hline
 \hline
 $ 5$ & $0.89$ & $0.98$ & $1.00$ & $0.98$ & $0.99$ & $1.00$ \\
 $ 9$ & $0.38$ & $0.90$ & $1.00$ & $0.10$ & $0.88$ & $0.97$ \\
 $15$ & $0.00$ & $0.56$ & $0.93$ & $0.00$ & $0.54$ & $0.94$ \\
\hline
 f$_\mathrm{g}$
 &  \multicolumn{3}{c |}{G1 ($c=9$)}
 &  \multicolumn{3}{c  }{G2 ($c=9$)} \\ 
\hline
\hline
 $0.1 $ & $0.99$ & $0.97$ & $1.00$ & $0.90$ & $0.99$ & $0.99$ \\
 $0.35$ & $0.38$ & $0.90$ & $1.00$ & $0.10$ & $0.88$ & $0.97$ \\
 $0.5 $ & $0.12$ & $0.89$ & $1.00$ & $0.00$ & $0.87$ & $0.97$ \\
\hline 
\hline
\end{tabular}
\end{center}
\label{tab:ezrH}
\end{table}%

\subsection{Comparison with previous works}

\begin{table*}[]
\caption{Mass and metal ejection efficiencies from various works.}
\begin{center}
\begin{tabular}{c c c c c c c c c }
\hline
\hline
Author & M$_\mathrm{SB}$ & E$_\mathrm{SB}$ & R$_\mathrm{c}$ & M$_\mathrm{g}$ 
       & $f_\mathrm{g}$  & c\tablenotemark{a} &  $\xi_\mathrm{m}$
       & $\xi_\mathrm{z}$ \\
 & ($\times 10^5$ M$_\odot$) & ($\times 10^{51}$ erg) & (pc) & ($\times 10^8$ M$_\odot$) & & & &  \\
\hline
\hline
\citetalias{1999ApJ...513..142M}
  & $0.1$ &3.78$\times$10$^4$& $100$ & $1$ & \nodata & $f$ & $0.000$ & $1.0$    \\
  & $1.0$ &3.78$\times$10$^5$& $100$ & $1$ & \nodata & $f$ & $0.001$ & $1.0$    \\
\citetalias{2004ApJ...617.1077F} 
            & \nodata        &1.5$\times$10$^3$& \nodata & $10$ & \nodata & \nodata & $0.023$ & $0.91$    \\
 $\dag$& \nodata        &1.5$\times$10$^3$& \nodata & $10$ & \nodata & \nodata & $0.032$ & $0.47$    \\
   & \nodata  &1.5$\times$10$^4$& \nodata & $10$ & \nodata & \nodata & $0.016$ & $0.99$    \\
\citetalias{2011RMxAA..47..113R} 
  & $0.55$  &2.08$\times$10$^4$& $100$ & $1.4$ & $0.35$ & $9$ & $0.500$ & $0.800$ \\
  & $5.55$  &2.08$\times$10$^5$& $100$ & $1.4$ & $0.35$ & $9$ & $0.950$ & $0.980$ \\
  & $27.5$  &1.04$\times$10$^6$& $100$ & $1.4$ & $0.35$ & $9$ & $0.990$ & $1.0  $ \\
  & $2.75$  &1.04$\times$10$^5$& $100$ & $4.7$ & $0.35$ & $9$ & $0.300$ & $0.800$ \\
  & $27.5$  &1.04$\times$10$^6$& $100$ & $4.7$ & $0.35$ & $9$ & $0.880$ & $0.980$ \\
  & $55.0$  &2.08$\times$10$^6$& $100$ & $4.7$ & $0.35$ & $9$ & $0.920$ & $1.0  $ \\
\citetalias{2013AA...551A..41R} 
  & \nodata   &\nodata& $200$ & $1$ & $0.6$ & $v$ & $0.606$ & $0.826$  \\
  & \nodata   &\nodata& $200$ & $1$ & $0.6$ & $f$ & $0.598$ & $0.872$  \\
  &  \nodata  &\nodata& $200$ & $1$ & $0.6$ & $s$ & $0.558$ & $0.599$  \\
This work 
  & $0.55$  &2.08$\times$10$^4$& $50$ & $1.4$ & $0.35$ & $9$ & $<0.01$ & $0.00$ \\
  & $5.55$  &2.08$\times$10$^5$& $50$ & $1.4$ & $0.35$ & $9$ & $0.23$ & $1.00$ \\
  & $27.5$  &1.04$\times$10$^6$& $50$ & $1.4$ & $0.35$ & $9$ & $0.72$ & $1.00$ \\
  & $2.75$  &1.04$\times$10$^5$& $50$ & $4.7$ & $0.35$ & $9$ & $<0.01$ & $<0.01$ \\
  & $27.5$  &1.04$\times$10$^6$& $50$ & $4.7$ & $0.35$ & $9$ & $0.06$ & $0.98$ \\
  & $55.0$  &2.08$\times$10$^6$& $50$ & $4.7$ & $0.35$ & $9$ & $0.29$ & $1.00 $ \\
\hline
\hline
\end{tabular}
\tablenotetext{a}{$f$=flattened disk, $v$=very flattened disk,
  $s$=spherical.\\ $\dag$ same as the model above,  with a starburst
  placed of centered (half-way in the galactic disk).
}
\end{center}
\label{tab:comp}
\end{table*}%

In Table \ref{tab:comp}  we have compiled various values of gas mass
ejection efficiency and metal ejection efficiency reported in
previous works. To facilitate the comparison, we chose for Table
\ref{tab:comp} models with comparable parameters to the
  simulations presented here, and we add a selection of the models in
  this work to the table.

Given the L${_\mathrm{m}}$ of the starburst and the flat distribution
of the gas mass of the galaxy (related to the parameter $c$), two models
of \citetalias{1999ApJ...513..142M} can be compared to our models of G1
and G2, with nuclear starburst ($c$=9 and the lowest
L${_\mathrm{m}}$, $\xi_\mathrm{m}$,  Table \ref{tab:comp}.
In the case of $\xi_\mathrm{m}$ the results of
\citetalias{1999ApJ...513..142M} and ours results are in 
good agreement, both studies is concluded that the such
  galaxies can not loose a significant amount of mass.
For $\xi_\mathrm{z}$ we found the opposite 
of what is presented by \citetalias{1999ApJ...513..142M}, in their
models the metal loss is 100\% while in our models the metal
loss is less than 1 percent.

We have 6 simulations in common with
\citetalias{2011RMxAA..47..113R}. The first three models of
\citetalias{2011RMxAA..47..113R} are equivalent
to our models for G1 with $c$=9 in Table 2 and 3 for $\xi_\mathrm{m}$
and  $\xi_\mathrm{z}$, respectively. And the following three models of
\citetalias{2011RMxAA..47..113R} correspond to  
our models for G2 with $c$=9 in Table 2 and 3 for
$\xi_\mathrm{m}$ and $\xi_\mathrm{z}$, respectively. 
The simulations with small L$_\mathrm{m}$ for both
galaxies differ considerably in both efficiencies, $\xi_\mathrm{m}$
and $\xi_\mathrm{z}$. The values in \citetalias{2011RMxAA..47..113R}
are high compared with those obtained in the present analysis. For
larger L$_\mathrm{m}$, the mass loss of \citetalias{2011RMxAA..47..113R} is greater than the mass loss found 
in the present work (30\% to 80\% higher), but the loss of metals is very similar,
close to 100\%.

\citetalias{2013AA...551A..41R} presented  models with different
galactic mass distributions, which we have named here as very flat
(denoted as $v$ in  the $c$ parameter column),
flat ($f$) and spherical ($s$).
G2 with the smaller L$_ \mathrm{m}$ and different values of $c$,
 where $c$=15 is comparable with the very flat model, $c$=9 to the
 flat ones, and $c$=5 to the spherical one.
The values of both $\xi_\mathrm{m}$ and $\xi_\mathrm{z}$ of
\citetalias{2013AA...551A..41R} are considerably different from ours,
the most important difference between their models and ours is
that they consider a continuous star formation, and a larger region
for the starburst.

In a similar study, \citet{2004ApJ...617.1077F}
presented a series of models that inject the energy from SN
explosions and estimate the efficiency mass and metals
(metal rich material imposed with the SN) loss. Their simulations have
some important differences, for instance, their models include a
continous SN rate, and the explosions are distributed both in time,
within the galactic disk and  their models do not include
a self-gravity. 
At the same time their galactic disks with concentrated bursts
are an order of magnitude more massive than ours. 
Their models m1, m2 and m5 can be more or less
compared with ours. Although different in mass and energy input, both
models are concentrated starbursts (all SNe are imposed at the same
galactic location, but still over a period of 
$50~\mathrm{Myr}$), but m2 is off-centered.
They found, see Table \ref{tab:comp}, a higher
efficiency in the nuclear burst (m1) and very similar mass and metal ejection
efficiencies when their m5 (more energetic nuclear burst) is compared
with our model with nuclear starburst  with mechanical energy
L$_m$=3$\times$10$^{39}$~erg/s, $f_g$=0.35 and $c$=9.
In their off-centered model m2 the mass and 
metal efficiencies are comparable with our results, but they find a
trend of lower metal loss eficiencies with increasing the
galactocentric position of the starburst. 
To test this, we have run a set of simulations of G1 with
L$_m$=3$\times$10$^{39}$~erg/s, $f_g$=0.35 and $c$=9, placing  the
starburst at $R=0$, $0.5$, $1$, $1.5$, and $2 R_0$.
We obtained mass ejection efficiencies of $0.01$, $0.02$,
$0.03$, and $0.04$  and  metal
ejection efficiencies of $0.0$, $0.25$, $0.38$, $0.24$, $0.24$, respectively.
We attribute the difference (high metal ejection for non-nuclear
starburst in comparison with the nuclear burst), to the flaring of
their disks. As the supershell formed by the SNe has to overcome a
larger pressure at larger galactocentric radii. In our models the
disks much less flared, and while the pressure is similar, the
galactic potential is shallower at larger radii, thus facilitating the
escape of material.

\section{Conclusions}\label{sec:conclusions}

We have developed an extensive set of numerical simulations of dwarf
galaxies with galactic winds. The galactic winds are produced by
supernova explosions for a given starburst. The studied galaxies have
masses of: 1.4$\times$10$^{8}$ M$_\odot$ and 4.7$\times$10$^{8}$
M$_\odot$, G1 and G2 respectively. For each galaxy three different
energies were considered and injected into the starburst.
Also, the position of the starburst on the disk was varied, as well as the
concentration parameter of the galaxy, and its gas mass fraction of
the disk.

From simulations, we analyze the mass and metals ejection efficiency,
and we found that: 

For different concentration parameters, we found that the metal
ejection efficiency (loss of the mass from the starburst
  itself) are very dependent on the parameter $c$ adopted. 
However, it is very similar irrespective of the difference in
mechanical  luminosities.

At the same time the mechanical luminosity does have an
  appreciable effect on the efficiency of mass (pristine ISM) loss,
  with high mechanical luminosities yielding higher losses.

 When the gas mass fraction of the disk varies in the simulations, in
general, the dependence of both efficiencies with the
luminosity, is very similar to the simulations where $c$ varies.

From our models we can say that the intense non-nuclear
  (off-centered) starbursts ($L\ge 15 \times 10^{39}\mathrm{erg~s^{-1}}$ for
  G1 and $L\ge 50 \times 10^{39}\mathrm{erg~s^{-1}}$ for G2) produce a
  metallic wind. This can be seen as the value of $\xi_\mathrm{z}$ increases to almost
  one for $R>0.5~R_0$, while the value of $\xi_\mathrm{m}$ tends to
  decrease as there is very little mass in the galaxy outskirts to be lost.
For the less energetic starbursts, the efficiencies depends on the
galaxy model (e.g. gas fraction or concentration index).


We found that most winds produced by starbursts in dwarf galaxies
have a high metal content (provided by the stars that form the wind),
and that these starbursts can not produce well mixed winds, because
the metals launched by the massive stars are unable to mix
  back into the galaxy efficiently.

\acknowledgments{We acknowledge support from CONACYT grants 167611 and
  167625 and the DGAPA-UNAM grants IA 103115, and IN 109715. We thank the anonymous referee for very relevant
comments that improved this paper.} \\

\bibliography{Master}

\end{document}